\documentclass{article}\usepackage{jkas2}
\runningauthor{Giovannini and Feretti}
\runningtitle{Radio Relics in Clusters of Galaxies}
\beginpage{1}
\endpage{5}

\begin{document}

\font\twelvei = cmmi10 scaled\magstep1 
       \font\teni = cmmi10 \font\seveni = cmmi7
\font\mbf = cmmib10 scaled\magstep1
       \font\mbfs = cmmib10 \font\mbfss = cmmib10 scaled 833
\font\msybf = cmbsy10 scaled\magstep1
       \font\msybfs = cmbsy10 \font\msybfss = cmbsy10 scaled 833
\textfont1 = \twelvei
       \scriptfont1 = \twelvei \scriptscriptfont1 = \teni
       \def\mit{\fam1 }
\textfont9 = \mbf
       \scriptfont9 = \mbfs \scriptscriptfont9 = \mbfss
       \def\bmit{\fam9 }
\textfont10 = \msybf
       \scriptfont10 = \msybfs \scriptscriptfont10 = \msybfss
       \def\bmsy{\fam10 }

\def\etal{{\it et al.~}}
\def\eg{{\it e.g.,~}}
\def\ie{{\it i.e.,~}}
\def\lsim{\raise0.3ex\hbox{$<$}\kern-0.75em{\lower0.65ex\hbox{$\sim$}}}
\def\gsim{\raise0.3ex\hbox{$>$}\kern-0.75em{\lower0.65ex\hbox{$\sim$}}}

\title{Radio Relics in Clusters of Galaxies}

\author{Gabriele Giovannini$^{1,2}$ and Luigina Feretti$^2$}
\address{$^1$ Dipartimento di Astronomia,
Universita' di Bologna, Italy \\
{\it E-mail: gabriele.giovannini@unibo.it}}
\address{$^2$ Istituto di Radioastronomia, IASF-CNR Bologna, Italy \\
{\it E-mail: lferetti@ira.cnr.it}}

\address{\normalsize{\it (Received October 31, 2004; Accepted December 1,2004)}}

\abstract{In this paper we review the observational results on Relic 
radio sources in 
clusters of galaxies. We discuss their observational properties, 
structures and radio spectra.
We will show that Relics can be divided according to their size, 
morphology,
and location in the galaxy cluster. These differences could be related
to physical properties of Relic sources.
The comparison with cluster conditions
suggests that Relics could be related to shock waves originated by
cluster mergers.}


\maketitle

\section {Introduction}

An increasing number of clusters of galaxies is known to contain large-scale 
diffuse
radio sources whose origin is not related to the activity of Active Galactic
Nuclei (AGN) in cluster galaxies, but to the  intracluster medium.
These sources are classified in two groups, radio halos and Relics, according
to their location with respect to the cluster center: radio halos show
a regular emission around the cluster center,
Relics are not located at the cluster center and in most cases are
in peripheral cluster regions and have an irregular, elongated shape
(see e.g. Giovannini \& Feretti 2002).

The existence of Relic sources reveal the presence of cluster wide magnetic 
fields of about 0.1 - 1 $\mu$G and of relativistic
electrons of $\sim$ Gev energies
in large regions where the number of galaxies is scarce
and the density of the hot Intra 
Cluster Medium (ICM) is low (peripheral regions).

Several studies of radio halos and their hosting clusters have been recently
published, thus our knowledge of their characteristics and physical properties
has  largely improved (see e.g. Feretti 2003, 2004). 
About radio Relics instead,  the available data
and knowledge are still poor and the number of well studied sources is limited.
Many Relics have not yet been studied or  observed in detail and deep
high resolution radio maps are still missing,
therefore their reported size
and radio flux density  could be under-estimated.
Most radio spectra have been obtained with a poor frequency coverage
and very few spectral index images have been published.
Moreover many Relics, being at the cluster periphery, 
are in the external regions or outside 
of the available X-ray images. In these 
cases a comparison
between radio and X-Ray data is difficult or not possible.

In this paper we will present the general properties of this
puzzling class of radio sources to improve our understanding
of the diffuse non thermal emission in clusters of galaxies. 

To estimate intrinsic parameters we use a cosmology
with H$_0$ = 70 km sec$^{-1}$ Mpc $^{-1}$, $\Omega_m$ = 0.3 and 
$\Omega_{\Lambda}$ = 0.7.

\section {General Properties}

We define a Relic as an extended diffuse synchrotron emission not located
at the cluster center and not identified with the activity of one or more
cluster galaxies. In most cases Relics show an elongated structure, a steep 
spectrum ($\alpha$ $>$ 1, assuming S$_\nu$ $\propto$ $\nu^{-\alpha}$),
and are linearly polarized at a level of $\sim$ 10 -- 30 \%.

Currently we know $\sim$ 30 clusters of galaxies where at least 
one Relic source is present. 
Most of them are rich Abell clusters, but a Relic source is present
in a X-Ray selected cluster (RXS J1314236-251521, 
Valtchanov et al. 2002),
and one belongs to a poor cluster (S0753, Subrahmanyan et al. 2003).

Kempner et al. (2004) suggested a physical classification scheme
for Relics, and distinguished three different types of Relic sources. 
Here, we prefer
to present the properties of Relics simply according to their
observational structures and location, without relating them to physically
different classes.
We can group known Relics as discussed in the following sub-sections.

\subsection{Classic Elongated Relics}

We include in this class elongated and peripheral diffuse radio sources.
The prototype source of this class is the Relic 1253+275 in the Coma cluster
(see Giovannini et al. 1991 and Fig. 1). Other similar Relics  
 are e.g. in A2255, A2744, A1367 and A115. We note that the Relic in A115
(Govoni et al. 2001a) apparently starts at the main cluster center and
 is elongated towards the cluster periphery. Present data
do not allow to understand if it really starts in the central cluster
region or if this peculiarity is due to projection effects.

Most of these Relics are quite extended, with 
the linear size ranging from 400 to 1500 kpc. 
They show an asymmetric transversal profile, with a sharp edge 
and a flatter spectral index in the external
side and a smooth edge in the side near to the cluster center. The radio
emission is usually polarized. 
These characteristics and the presence in a few clusters of double
Relic sources (see Sect. III)
have suggested the link to models where cluster mergers produce
shock waves that propagate in both directions along the line that connects
the centres of the initial pre-merger clusters. 
These merger shock waves can revive 
fossil radio plasma present in cluster regions or to accelerate electrons
from the thermal plasma (Kempner et al. 2004).

\begin {figure}[t]
\vskip 0cm
\centerline{\epsfysize=8cm\epsfbox{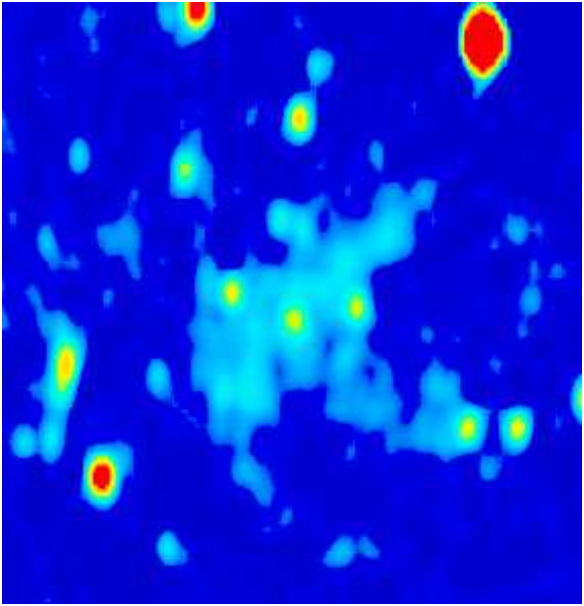}}
\vskip -0.2cm
\label{fig1}
\caption{Colour image of the Relic source 1253+275 in the Coma cluster observed
with the VLA at 90 cm (Giovannini et al. 1991). 
The strong point like source on the top is 3C 277.3 
(Coma-A)}
\vskip -0.5cm
\end{figure}

\subsection{Circular Peripheral Relics}

Two clusters (A548b and A1664) host a Relic radio source 
which is clearly at the
cluster periphery, but shows an extended mostly circular shape
(Feretti et al. in preparation and Govoni et al. 2001a). 
In Fig. 2 we report the superposition 
of  X-Ray and NVSS (Condon et al. 1998) radio data for A1664. 
This peculiar morphology could suggest that 
the  elongated Relics might be  extended disk-like
Relics seen in projection. However, this is not the case because of the
too large number of Relics with 
elongated structure. Moreover, we note that the Relic in 
A1664 when observed at higher resolution shows evident substructures
(Govoni et al. 2001a).

\begin {figure}[t]
\vskip 0cm
\centerline{\epsfysize=8.cm\epsfbox{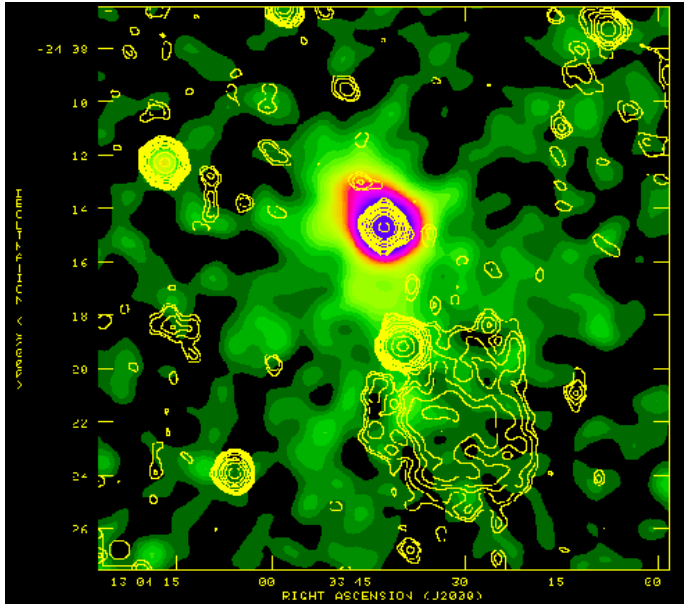}}
\vskip -0.2cm
\label{fig2}
\caption{Radio image (contours) of the radio Relic in the cluster A1664
(Govoni et al. 2001a)
from NVSS data, overimposed to the cluster ROSAT X-ray image  (colour)}
\vskip -0.5cm
\end{figure}

\subsection{Relic sources near the first ranked galaxy}

In some clusters  an extended diffuse emission is
located near the central First Ranked Galaxy (FRG, usually a cD galaxy), 
but not coincident with it. We classify these sources as Relics even if 
they are not located at the cluster periphery, since their connection to the 
activity of the FRG
is not clear (but see Fujita et al., 2002 for A133). 
The distances of these Relic sources from the cluster center are in the range
from $\sim$ 50 to $\sim$ 350 kpc.
We note that if these diffuse sources were old lobes of a previous activity
of the central galaxy we should expect to see in  most cases an almost
symmetric double
structure centered on the galaxy,
 whereas  in all these clusters only one diffuse emission is present.

These sources can be very small
($<$ 100 kpc in A133, A2063, A4038) but also quite large: the Relic in A85 is $\sim$ 360 
kpc
in size and that in A13 is $\sim$ 650 kpc.
They have a very steep spectrum ($\alpha$ $>$ 1.5 -- 2.0)
with evidence of high frequency 
steepening. The radio emission is strongly polarized and shows a regular
shape with evident filaments inside (see Slee et al. 2001 and Fig. 3). 

\begin {figure}[t]
\vskip 0cm
\centerline{\epsfysize=9cm\epsfbox{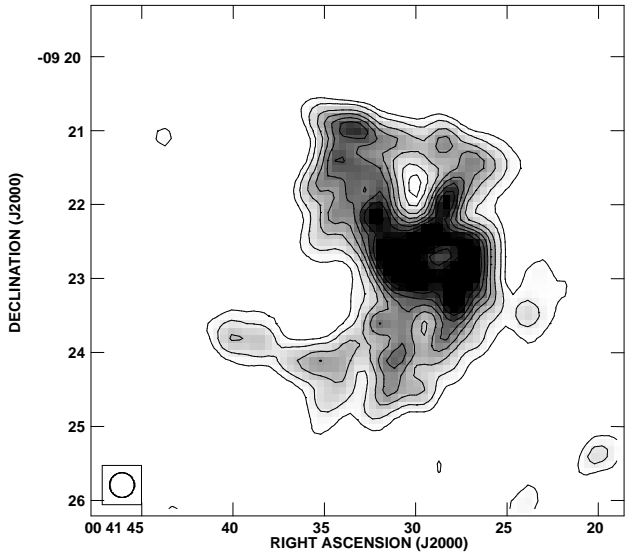}}
\vskip -0.5cm
\label{fig3}
\caption{Relic source in the cluster A85 from VLA radio data at 90 cm
(Giovannini and Feretti, 2000)}
\vskip -0.5cm
\end{figure}

\subsection {Relics at large distance from the cluster center}

In this sub-class we include two Relic radio sources very far
from the cluster center: 0917+75 (Fig. 4) tentatively identified with A786 and
the Relic probably associated with A2069.
 
0917+75 is an extended elongated diffuse emission (Dewdney et al. 
1991, Harris et al. 1993, Giovannini and Feretti 2000) located at 
$\sim$ 3.8 Mpc from the nearest rich cluster of galaxies; the Relic source in 
A2069
(Giovannini et al. 1999) is at $\sim$ 4.6 Mpc from the cluster center.
 
These distances are so huge that a connection 
between the Relic radio emission and the cluster is not straightforward. 
In these cases
the radio activity could be related to a possible filamentary structure
in a super-cluster structure (see next sub-section).

\begin {figure}[t]
\vskip 0cm
\centerline{\epsfysize=8.cm\epsfbox{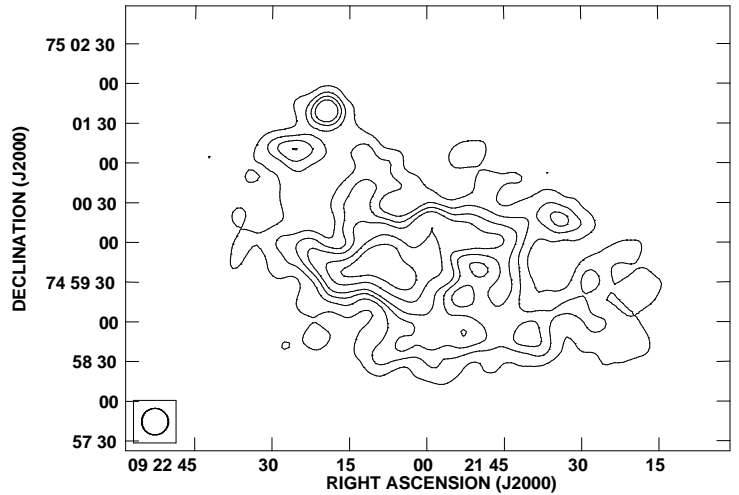}}
\vskip -0.5cm
\label{fig4}
\caption{Relic source 0917+75 tentatively associated with A786.
The image is from VLA data at 1.4 GHz with a HPBW = 20''(Giovannini and 
Feretti, 2000)}
\vskip -0.5cm
\end{figure}

\subsection {Filaments}

Radio emission has been detected in some cases 
from regions in between clusters of galaxies (filaments). 
Probably, these features
should not be considered Relic sources, because they 
may have a different origin and evolution. 

% given the existence of
%Relic sources very far from the cluster center and the bridge of radio
%emission found in between some 
%halo and Relic sources, we want here briefly to discuss 

Relic sources are detected in cluster peripheral regions. Can we
expect the presence of radio emission in even less dense regions as
filaments in between rich cluster of galaxies? 
A suggestion of a possible radio emission in filaments comes from the bridge
of radio emission visible in the region between Relics and halos in a few
clusters as Coma, A2255, and A2744.
In particular, in the Coma cluster a bridge of radio emission about 
1 Mpc in size is present between the central halo Coma C and the
peripheral Relic source 1253+275 (Kim et al. 1989, Giovannini et al. 1990). 
We note that this elongated radio emission is
oriented in the same direction as the Coma-A1367 supercluster. 
Also the two Relics at large distance from the cluster center (see previous
Sub-Section) could be related to the radio activity in a filament region.

Moreover, Bagchi et al. (2002) found a radio emission coincident with the
filament of galaxies ZwCl 2341.1+0000, 2.5 Mpc in size.
The possible detection
of radio emission in filaments would imply the existence of faint
magnetic fields in super-cluster regions and will raise questions about the 
origin of relativistic particles. 

\section {Clusters with Multiple Diffuse Sources}

\subsection{Double Relics}

In 6 clusters of galaxies,  radio images show 
two Relic radio sources located in the peripheral regions,
and symmetric 
with respect to the cluster center 
(see e.g. A2345 in Fig. 5; Giovannini et al. 1999).
In most cases these Relic sources show a classical  elongated 
structure. 
In the parent clusters, no  central extended halo source has been 
detected.
The prototype and best studied cluster with two symmetric Relics is A3667
(Rottgering et al. 1997, Johnston-Hollitt et al. 2002, 2003).

These structures suggest that Relics are related to the presence of shock
waves originated by mergers between clusters with approximately equal masses.
In this scenario, it is expected that 
 Relics should often come in pairs and be located on opposite
sides of the cluster along the axis merger, with the extended radio
structures elongated perpendicularly to this axis.

\begin {figure}[t]
\vskip 0cm
\centerline{\epsfysize=7.5 cm\epsfbox{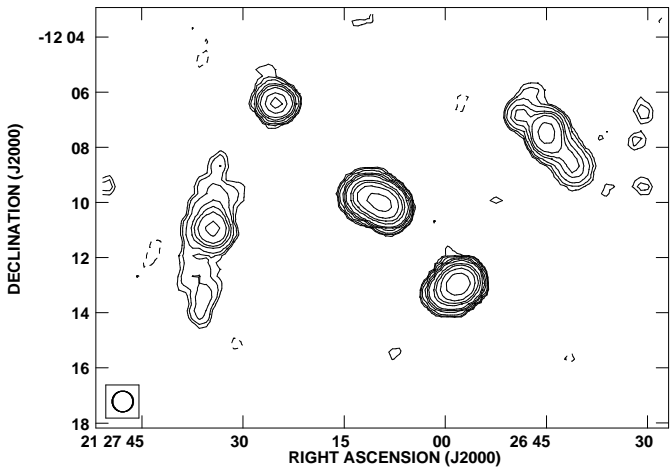}}
\vskip -0.2cm
\label{fig5}
\caption{Double Relics in the cluster A2345. Radio image is from the NVSS
survey at 1.4 GHz; see Giovannini et al. (1999).}
\vskip -0.5cm
\end{figure}

\subsection{Halo plus Relic}
 
In 7 cases, a cluster shows both a Relic and a halo radio emission. 
In most cases the 
Relic emission is elongated and at the cluster  periphery and it can be 
connected
to the halo source by a bridge of radio emission (e.g. Coma cluster, A2255 and 
A2744, see Fig. 6), but complex structures can also be present as in A2256.

The connection between halo sources and cluster mergers 
has been discussed in many papers (see e.g. Feretti 2004 and
references therein). 
A recent major  merger process is suggested to supply the energy necessary to 
reaccelerate the relativistic electrons in the central radio halo.
The same merger could be the origin of shock waves which supply the energy
to the peripheral
Relics. In this scenario, however, the origin 
of the bridge of radio emission connecting the halo to the Relic source
in some clusters has still to be clarified.

\begin {figure}[t]
\vskip 0cm
\centerline{\epsfysize=9.5cm\epsfbox{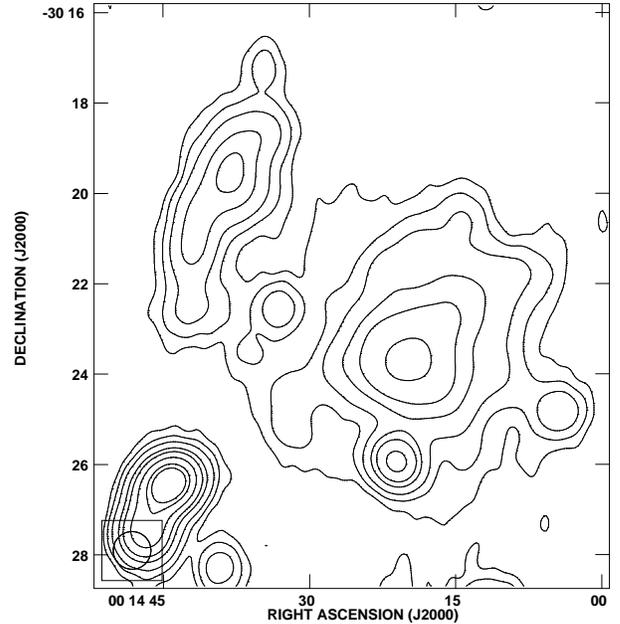}}
\vskip -0.2cm
\label{fig6}
\caption{Halo and Relic source in the cluster A2744. The image is from VLA
data at 1.4 GHz; see Govoni et al. (2001a)}
\vskip -0.5cm
\end{figure}

\section{Discussion}

\subsection{Redshift Distribution}

In fig. 7 we show the redshift distribution of clusters of galaxies where
at least a Relic source has been found. We could have missed Relic sources
at very low redshift because nearby Relics could have a too large angular 
extension to be detected in interferometric observations, due to  
missing short
uv spacings. However, thanks to the irregular and elongated structure
of most Relics,  this observational problem is not as strong
as in the more regular and diffuse halo sources.
In Fig. 8 we plot the Relic radio power
at different redshifts. The selection effect of the radio power with the
distance is well visible. If we consider only Relics
with a radio power larger than 10$^{24}$ W/Hz (visible at all distances)
we can conclude that the Relic distribution is homogeneous up to z $\sim$ 0.3.

\subsection {Radio Spectra}

Important information on the evolution and properties 
of Relics can be derived from
radio spectra, which reflect the energy distribution of relativistic
electrons.

The total radio spectra of Relic sources are steep ($\alpha$ $>$ 1). 
These are typical spectra of old radio
sources where the
radiative lifetime of relativistic electrons,
taking into account radiative and Inverse Compton losses, is of the order of
10$^8$ years. 
All small size Relics near the cluster brightest galaxy have very 
steep ($>$ 2) and curved radio 
spectra (see Slee et al. 2001). On the contrary, the Relic in 
the Coma cluster  and that in A786 
show a straight moderately steep ($\sim$ 1.2) radio spectrum. The number of 
Relics with accurate flux density measurements at different frequencies is too 
small to derive
useful considerations. 

The spectral index distribution
is known only in a few sources as A3667 (Rottgering et al. 1997), 1253+275
in the Coma cluster (Giovannini et al. 1991)
and S0753 (Subrahmanyan et al. 2003). The external side, i.e. 
the more distant side
from the cluster center, is always sharper and characterized 
by a flatter spectrum, consistent with the presence of electron
reacceleration in 
an expanding merger shock, but more data on more Relics are necessary
to investigate this point.

\subsection{Models}

En{\ss}lin and Gopal-Krishna (2001) proposed a scenario 
where fossil relativistic
particles from no more visible radio sources can gain energy during 
adiabatic compression by shock waves produced by a merger event. 
Hoeft et al. 
(2004) found that cluster-wide shock fronts can revive old radio ghosts
when the thermal pressure is much higher than the magnetic field 
pressure. In this model a much higher occurence of radio Relics in peripheral
locations is expected because in the cluster center the radio plasma
ages faster (higher magnetic field) and shock waves are weaker.
In the hotter cluster center we have weak shock waves which steepen when they 
pass the cooler outer regions of the cluster where the compression shock 
factor increases (Hoeft et al. 2004). 

Alternatively, Relic radio emission could
be due to electrons direcly accelerated from the thermal plasma in shocks
(Kempner et al. 2004). This mechanism could explain double Relics and
very extended elongated sources, which are more difficult
to relate to the fossil electrons supplied by single radio sources. 

Tests of these models are difficult because of the very low X-ray brightness
of the peripheral cluster regions. Data are presently available for
A754, where the radio Relic (Kassim et al. 2001,
Bacchi et al. 2003) is found to be at the same position of the hot region
found by Markevitch et al. (2003), which indicates the presence of a shock 
wave.

Finally we need to
understand why in many clusters with a merger activity, we do not see Relics.

\begin {figure}[t]
\vskip -1.cm
\centerline{\epsfysize=10.cm\epsfbox{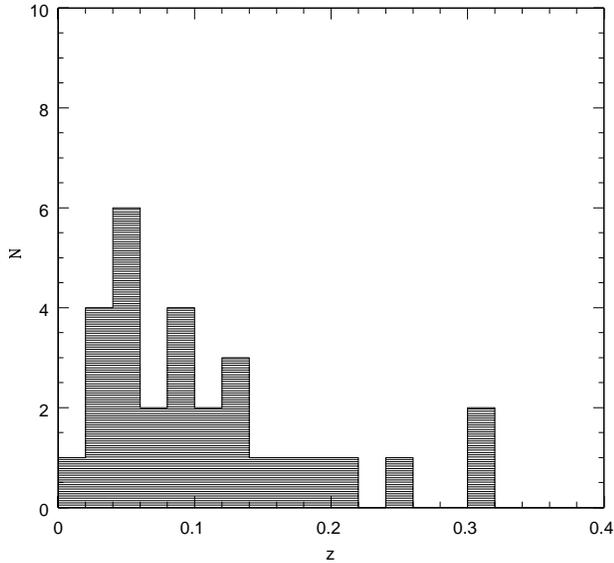}}
\vskip -0.2cm
\label{fig7}
\caption{Distribution in redshift (z) of clusters with at least a Relic
radio source}
\vskip -0.5cm
\end{figure}

\begin {figure}[t]
\vskip -1.cm
\centerline{\epsfysize=10.cm\epsfbox{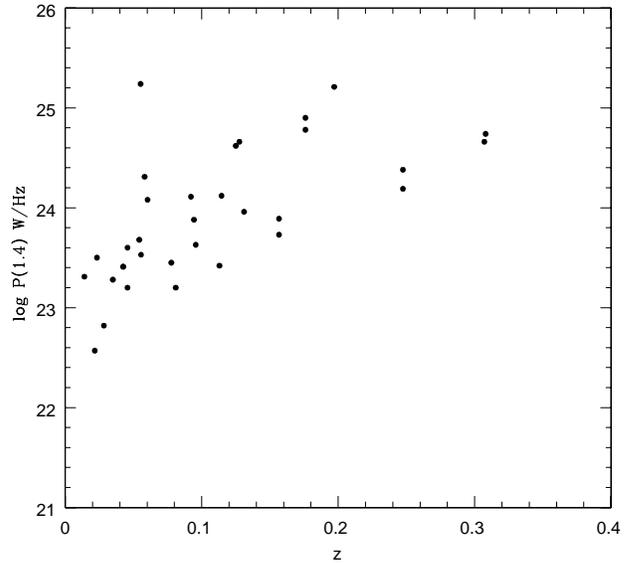}}
\vskip -0.2cm
\label{fig8}
\caption{Relic radio power at 1.4 Ghz in W/Hz versus cluster redshift (z)}
\vskip -0.5cm
\end{figure}

\subsection {Radio -- X-Ray comparison}

Studies of several radio halos permeating the cluster centers
have been recently published improving
our knowledge of this class of radio sources. In particular
it has been found that the radio power correlates with the cluster
X-ray luminosity (Bacchi et al. 2003) and that in well resolved clusters a 
point-to-point spatial correlation is observed between the X-ray and
radio brightness (Govoni et al. 2001b).

Relics are mostly in
peripheral regions, thus we do not expect similar strong correlations
to be present between relativistic electrons and hot gas,
however a comparison between Relic and X-Ray is important to test
the energy origin of Relic radio sources.

For this reason we have compared the Relic total radio power at 1.4 GHz with 
the parent cluster X-Ray luminosity. Despite of the low radio data quality
for some Relic sources, the data show a clear correlation: 

P$_{1.4 GHz}$ $\propto$ 10$^{K}$ L$_{X bol}$

where the slope K is in the range from 0.8 to 2.2.
We note that this slope
is in agreement with the slope (1.68) found by Bacchi et al. (2003) for 
the central halo
sources, although there is a much larger dispersion.
The large dispersion of the data may be due to the lack of
homogeneous and deep radio data for many Relic sources, but
may also indicate that the connection between the thermal and
relativistic plasma in Relics is weaker than in halos.
The existence of a correlation is nevertheless very important
and could be explained by the link between Relics and cluster
mergers.

To support this result we note that all clusters of galaxies which
exhibit a Relic source, and  which 
have been analyzed  in detail in the optical and/or X-Ray domain,
are  characterized by the presence  
of merger events (e.g. A754, Coma, A2256, A3667, etc.).
In a few clusters (e.g. A85, A115, A133), also a cooling core
is present, probably because the merger event 
is not strong enough (off-axis merger, or merger between 
sub-clusters of very different masses) to destroy the central 
cluster cooling flow.
 Therefore, also a merger event which is
not able to strongly influence the cluster center
(in these cases no halo source is expected and the
cooling flow is not destroyed)
seems to be able to reaccelerate particles in cluster peripheral regions
and to give origin to peripheral Relic sources.

\section {Conclusions}

1) The number of known Relics is increasing, however, better data are necessary
for many of them to allow a proper study.

2) Different Relic morphologies and distances from the cluster center could
be correlated with different Relic properties and origin. Elongated Relics
are located in peripheral regions, are polarized and show straight spectra.
Relics located near a FRG
have smaller size and very steep curved spectra. 
They show a filamentary polarized structure. In some clusters two Relics
located symmetrically with respect to the cluster center are present.
These structures suggest that Relics are tracers of shock waves.
The presence of a halo and a Relic radio source in the same cluster
confirms the correlation between cluster mergers and the presence of diffuse 
emission.

3) A correlation is found  
between the Relic radio power and the cluster X-Ray luminosity.
Although it shows a large dispersion, it is 
in agreement with the correlation found for halo radio sources.

4) Some Relics are present in clusters showing a central cooling core.
This suggests that Relics may originate not only from strong mergers,
but also from  minor or
off-axis mergers, which do not destroy the cooling flow.

5) The presence of radio emission in extended filaments is 
proved by the radio emission
in the Coma cluster and in the galaxy filament ZwCl 2341.1 +0000.

6) The next generation of radio telescopes (LOFAR, LWA and SKA) is necessary
to properly study Relics and filaments.

\acknowledgements{It is a pleasure to thank Hyesung Kang and Dongsu Ryu
for the invitation to such an interesting conference and for the warm
hospitality. We want also to thank them and the local people for the
very good organization.}

\end{document}